\documentclass{kluwer}    
\usepackage{graphicx}
\begin{document}                                                                                   
\begin{article}
\begin{opening}         
\title{CG~J1720-67.8: Radio and Integral Field Optical
Observations\thanks{Based on data collected at the Anglo-Australian Observatory, proposal
AAT/02A/24, and at the Australia Telescope Compact Array, proposal C1026.}} 
\author{Sonia \surname{Temporin}\thanks{Supported by the Austrian Science Fund 
(FWF) under project P15065}\email{giovanna.temporin@uibk.ac.at}} 
\institute{Institut f\"ur Astrophysik, Leopold-Franzens Universit\"at Innsbruck,
Austria}
\author{Lister \surname{Staveley-Smith}\email{Lister.Staveley-Smith@csiro.au}}
\institute{Australia Telescope National Facility, CSIRO, Australia} 
\runningauthor{S. Temporin and L. Staveley-Smith}
\runningtitle{CG~J1720-67.8: Radio and Integral Field Optical Observations}

\begin{abstract}
Our previous studies of the ultracompact galaxy group CG~J1720-67.8 have
revealed properties, which suggest that the group is in a very advanced 
evolutionary state.
We present here new observations in the radio and optical regimes, which have 
been obtained in  order to further investigate the dynamical and evolutionary 
state of the group.
Velocity fields reconstructed from integral field spectra of two regions of 
the group, obtained with SPIRAL at the Anglo-Australian Telescope (AAT), 
show some degree of distortion and a considerable velocity gradient across 
one of the TDG candidates. Estimates of the HI content and of the overall 
star formation rate of the group are obtained from radio observations with 
the Australia Telescope Compact Array (ATCA) in the 21 cm line and in the 
20 cm continuum.
\end{abstract}

\end{opening}           

\section{Introduction} 

CG~J1720-67.8 (z = 0.045) is an ultracompact galaxy group \cite{wtk99},
whose members have a median projected separation
of 6.9 kpc (H$_0$ = 75 km s$^{-1}$ Mpc$^{-1}$), a line-of-sight velocity
dispersion of $\sim$ 65 km s$^{-1}$, and show strong signs of mutual
interactions. Our previous studies \cite{te03,te02,TFVA02} suggest that 
this group is very evolved and therefore it offers us the rare chance to study
the final evolutionary phases of compact groups (CGs).
We give below a brief description of the group's components, which are labeled
in Fig.~1 (top-left) onto a R-band image obtained at the ESO 3.6 m telescope.

\emph{Galaxy 1:} A blue (B$-$V = 0.47) starburst galaxy with an exponential 
bulge and a bulge-to-total light ratio B/T $\sim$ 0.4, whose general 
properties are consistent with an Sc type. Since no spiral arms are visible,
the galaxy could have lost its outermost layers in the interaction. A bridge
of matter apparently connects this galaxy to galaxy 2.

\emph{Galaxy 2:} An S0 galaxy (B$-$V = 0.88) with a de Vaucouleurs bulge, which
dominates its light (B/T $\sim$ 0.7), and traces of central star formation.
Evolutionary synthesis models \cite{TFVA02} suggest that this galaxy might be
the result of a $\sim$ 1 Gyr old merger.

\emph{Galaxy 4:} A disk dominated (B/T $\sim$ 0.2), moderately blue (B$-$V =
0.55), starburst galaxy with an exponential bulge. No obvious spiral arms are
visible, but the galaxy shows internal structures, is in close interaction
with galaxy 2, and appears connected to object 3+9, at the base of an 
outstanding tidal structure (``arc'').

\emph{Objects 3+9 and 7(+8):} These objects are actively starforming, blue 
condensations (B$-$V $\sim$ 0.3), a few kpc in size, located at the opposite 
tips of the group's tidal arc. They are promising tidal dwarf galaxy (TDG)
candidates \cite{te02}.

\emph{Objects 10 and 12:} These are moderately blue, less concentrated 
structures located in the central part of the tidal arc. Star formation 
activity is spectroscopically confirmed for object 10. These could be 
complexes of giant H~II regions or TDGs in the process of formation.

\emph{Object 11:} This knot is embedded in a ring-like structure in the
group's optical halo. Is it maybe a vestige of a faded tail formed in the
merging process of galaxy 2?

\section{Dynamics and Kinematics: AAT-SPIRAL Observations}

Integral field spectra in the range 650 - 710 nm 
have been obtained in June 2002 at the AAT with SPIRAL for two 
9$^{\prime\prime}$.8$\times$10$^{\prime\prime}$.5
regions with the 14$\times$15 microlens array positioned as in Fig.~1, top-left panel.
Each lens imaged 0$^{\prime\prime}$.7$\times$0$^{\prime\prime}$.7 of the source. 
The images, reconstructed by
integrating the flux between 680 and 690 nm (H$\alpha$+continuum), are shown in
Fig.~1 (top and bottom central panels) after magnification and projection onto a
60$\times$56 pixel grid. The bridges of matter connecting 
galaxies 1, 2, and 4 are
evident. Radial velocities were measured in every spectrum, where the H$\alpha$
emission line was detected. The reconstructed velocity fields are shown in 
Fig.~1 (right-hand panels). Galaxy 1 shows a distorted velocity pattern, although the
observed velocity gradient across the galaxy is very small ($\sim$ 50 km s$^{-1}$).
Galaxy 2, with a velocity gradient of $\sim$ 120 km s$^{-1}$  in NE-SW direction,
shows a counter-rotation with respect to galaxy 1.
The bridge of ionized gas between the two galaxies shows an intermediate radial velocity, as
expected in case of a flow of matter. A considerable velocity gradient ($\sim$ 200 km s$^{-1}$)
is found across the TDG candidate 3+9. Such a gradient, especially at the location of this object, 
i.e. at the base of the tidal tail, could be a consequence of projection effects or 
streaming motion along the tail, but it could also indicate the presence of
a rotational motion. 

A complete coverage of the group with integral field spectroscopy
is still needed for a complete view of the complicated group dynamics.

\begin{figure}[H]
\centerline{
\vbox{
\hbox{
\includegraphics[width=4cm]{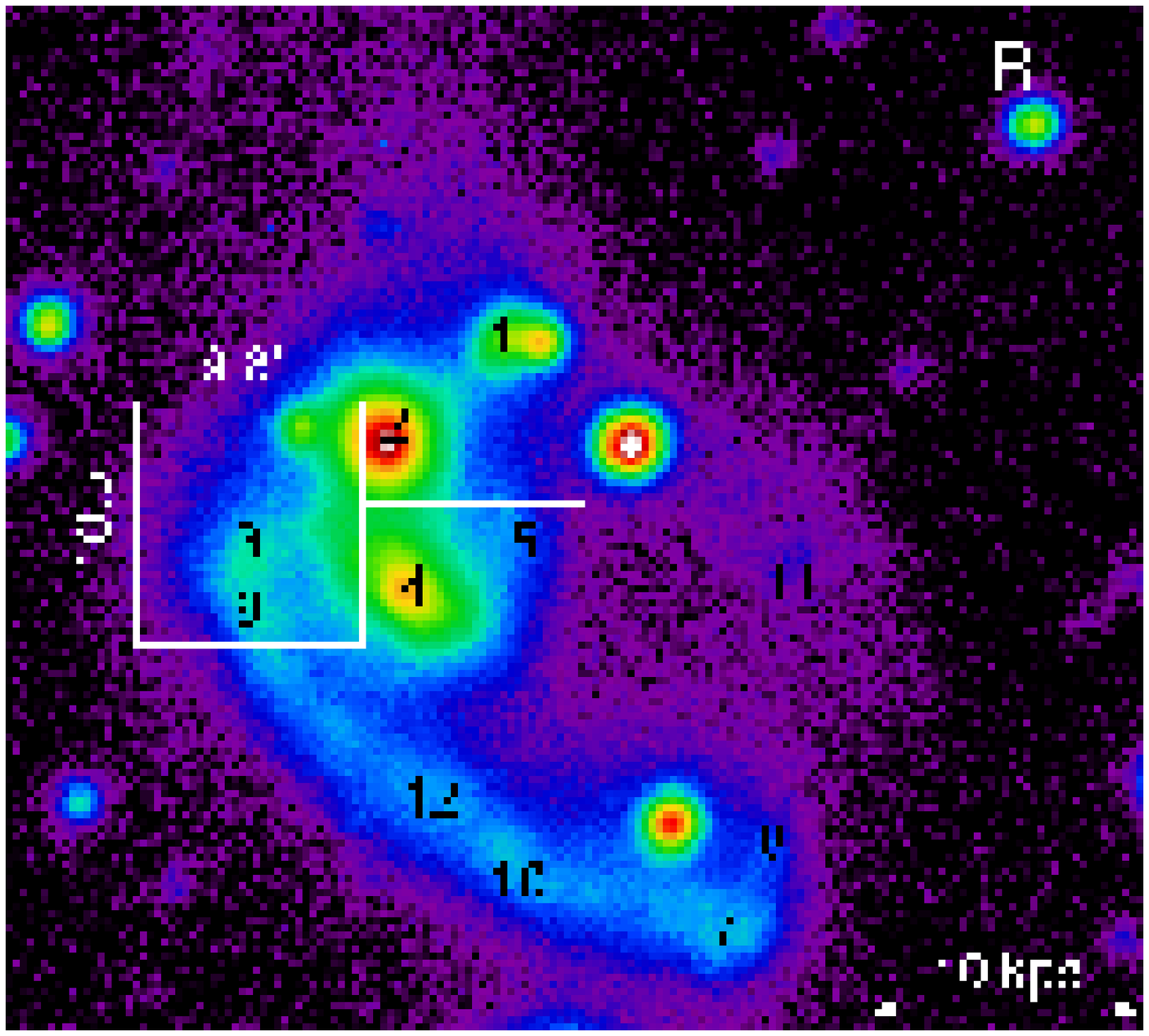}
\includegraphics[width=4cm]{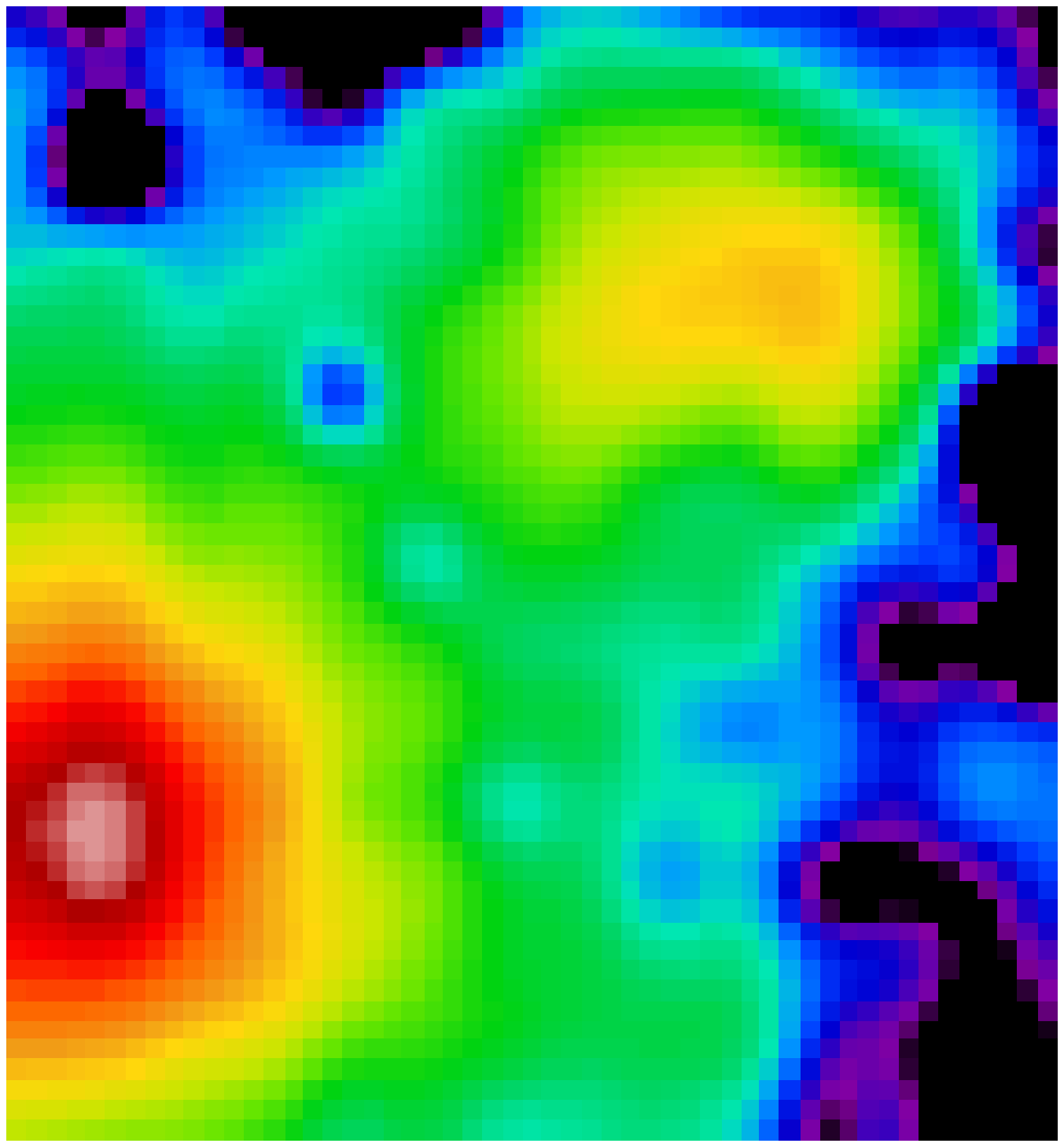} 
\includegraphics[width=4cm]{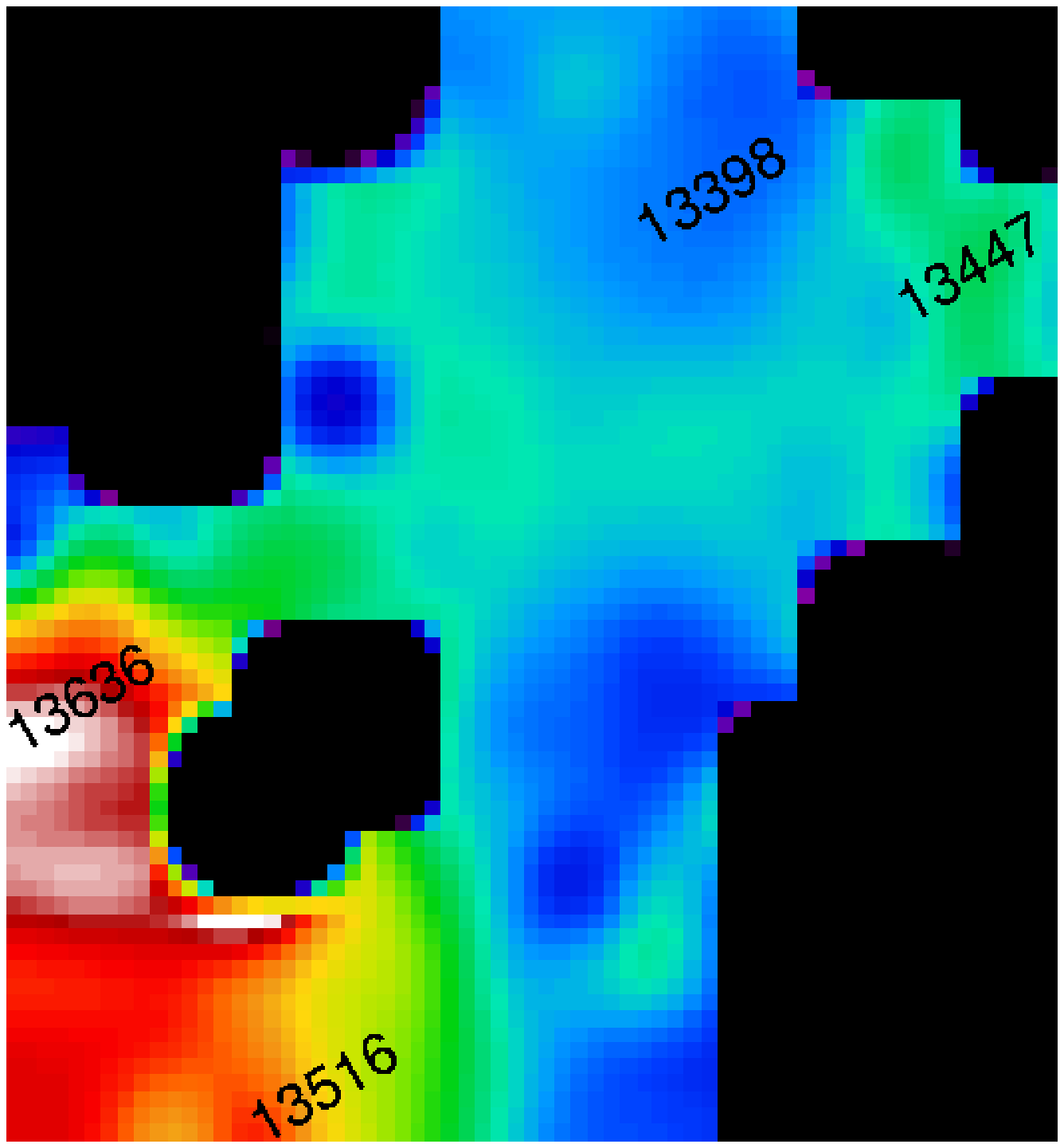}}
\hbox{
\includegraphics[width=4cm]{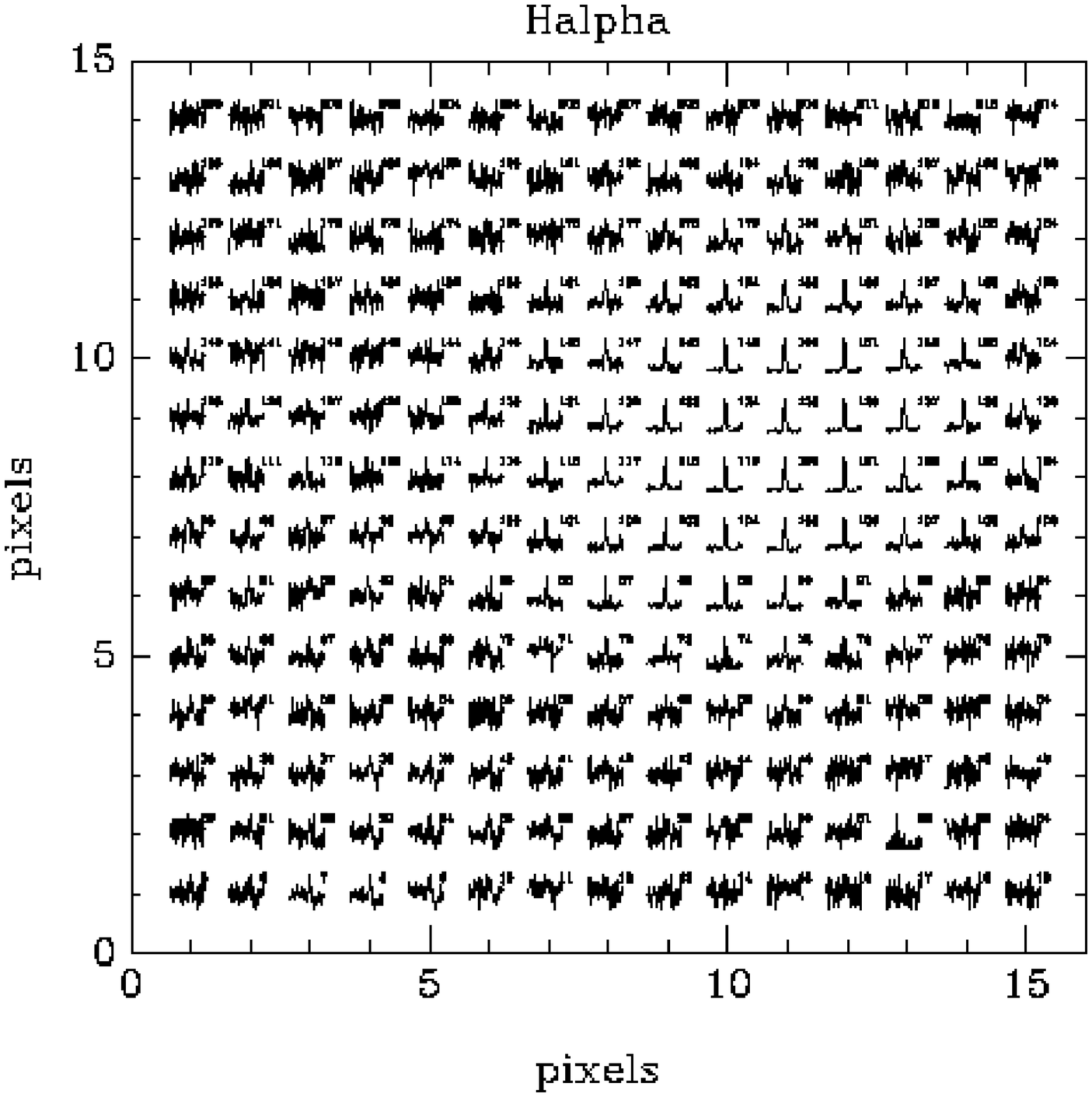}
\includegraphics[width=4cm]{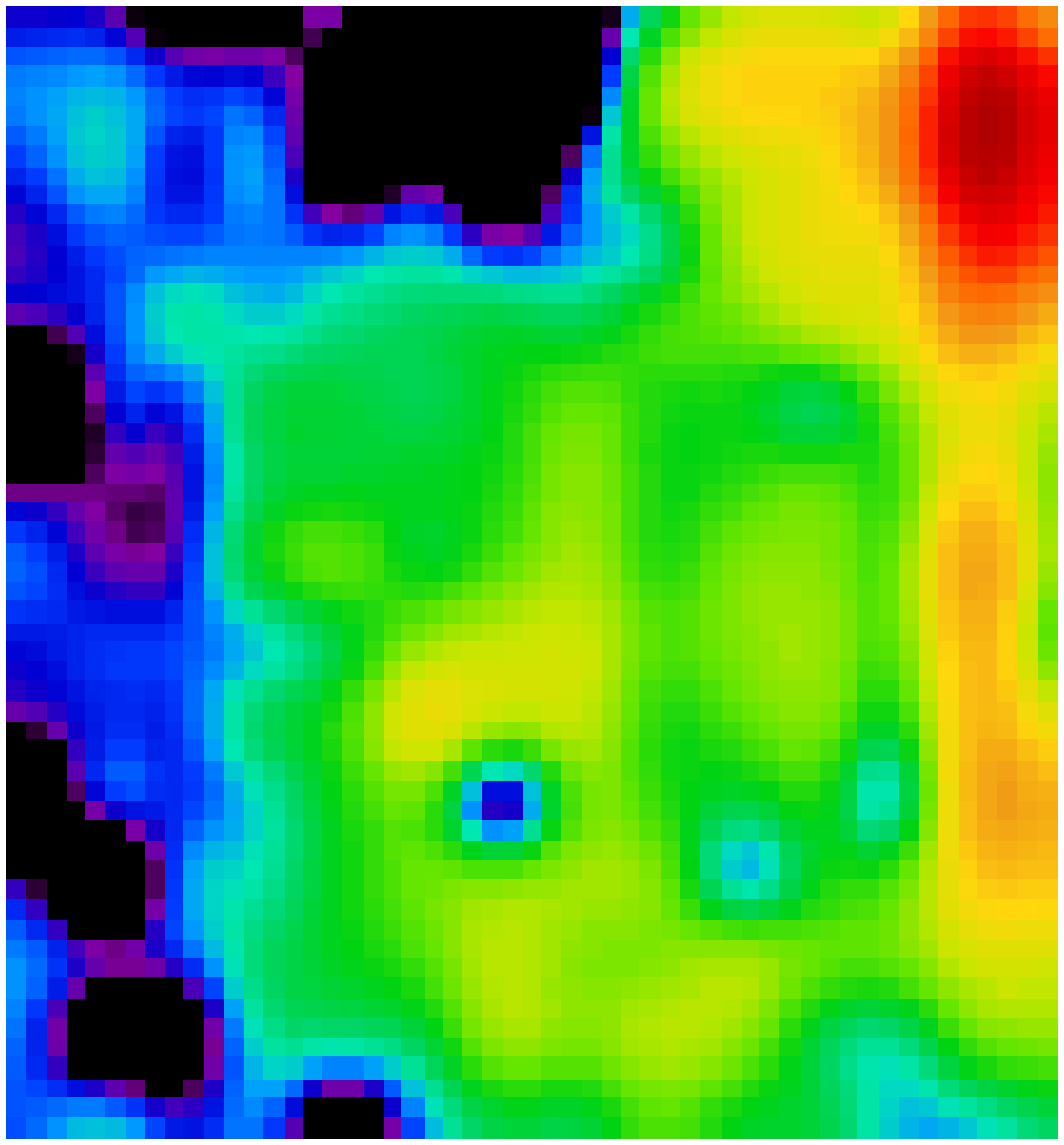}
\includegraphics[width=4cm]{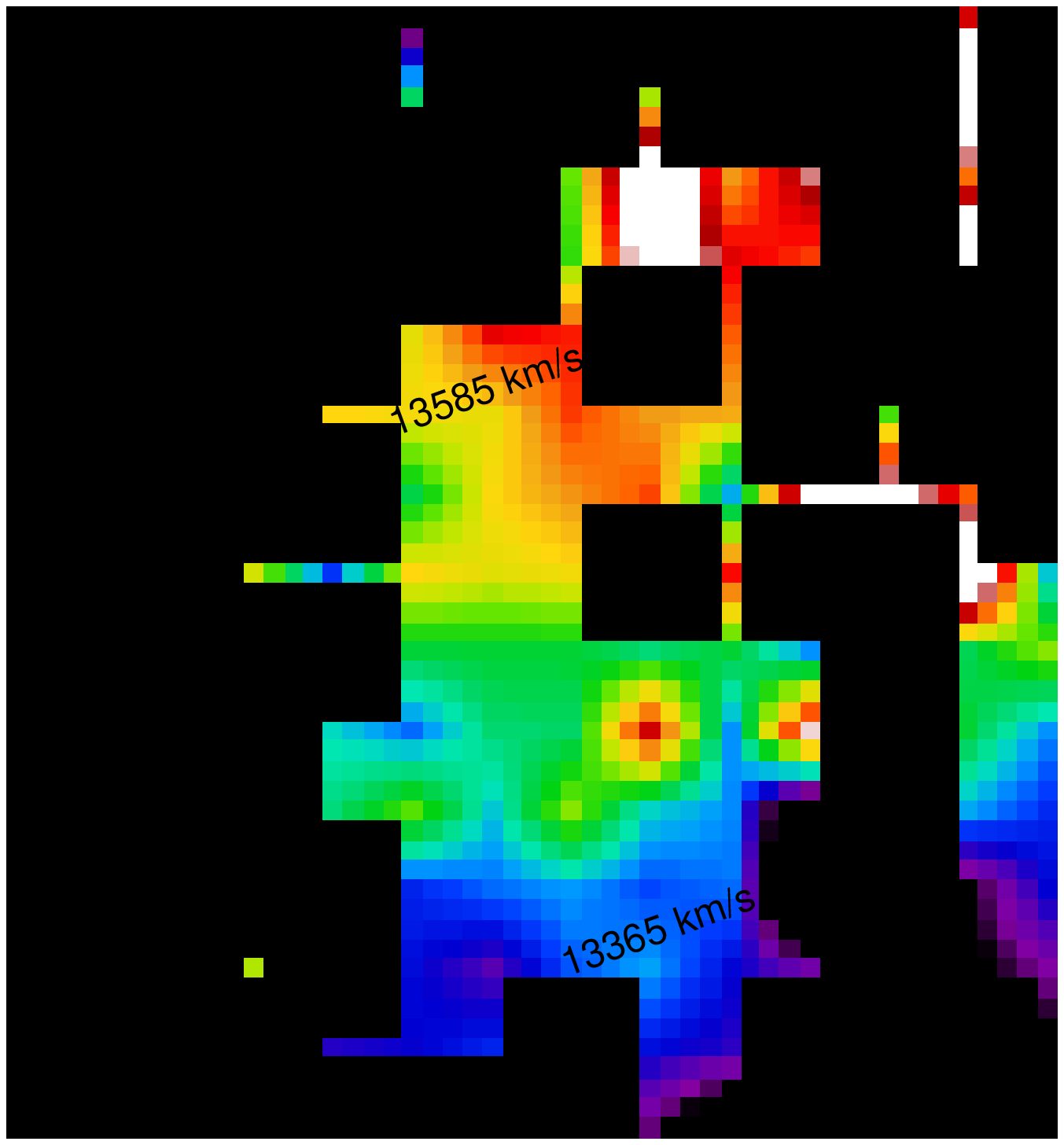}}}}
\caption{\emph{Top:} The members of CG~J1720-67.8 are labeled on a R-band
image from the ESO 3.6 m telescope. Two boxes indicate the two positions of the
SPIRAL array. In the central and right-hand panels the reconstructed H$\alpha$+continuum
 map and velocity field of galaxies 1 and 2 are shown after magnification and projection
 onto a 60$\times$56 pixel grid. \emph{Bottom-left:} Array of integral field spectra 
 of galaxies 1 and 2. \emph{Bottom-center and right:} Reconstructed, magnified, H$\alpha$+continuum
 map and velocity field of the TDG-candidate 3+9.\label{maps}}
\end{figure}

\section{Evolutionary State: HI Content of the Group}

The galaxy group, previously undetected in the radio regime, has been observed
with ATCA in January and February 2002 in the two configurations 
750A and 1.5A at the central frequency 1360 MHz with a bandwidth of 8 MHz. 
After 2$\times$12 hours synthesis (synthesized beam fwhm $\sim$ 20$^{\prime\prime}$)
the group still remained undetected in the 21 cm line. 
This gives an upper limit to the integrated 
HI mass of a few 10$^9$ M$_{\odot}$, which suggests that the group is
HI-deficient, consistent with it being in an advanced evolutionary state
\cite{vm01}. Instead an extended source with a deconvolved size of 
15$^{\prime\prime}.6\times9^{\prime\prime}.2$ and an integrated
flux of 3.6 mJy, centered on galaxy 4, was detected in the 20 cm
continuum. 
A high resolution map (beam fwhm $\sim$ 6$^{\prime\prime}$) was obtained by 
combining the wide-band continuum data 
with the two ATCA configurations and including data from the 6 km antenna.
The result, overlapped with the optical image of the group, is shown in Fig.~2.
The radio emission approximately follows the optical morphology of the
group and has a secondary peak in correspondence of object 7, one of the
most promising TDG candidates. Following Haarsma et al. (2000) we derived an overall 
star formation rate of $\sim$ 17 M$_{\odot}$ yr$^{-1}$.
\\

The new data presented here provide further confirmation of the advanced evolutionary
state of CG~J1720-67.8 and give additional evidence of the dynamical complexity of
this strongly interacting system.

\begin{figure}
\centerline{\includegraphics[width=6cm]{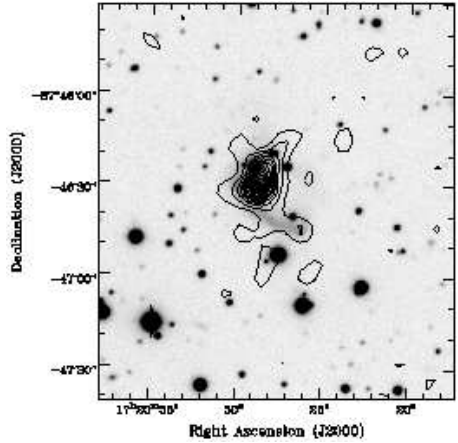}}
\caption{High-resolution map of the 20 cm radio continuum overlapped
with the optical image of the group.\label{radio_cont}}
\end{figure}

\end{article}
\end{document}